\documentclass[twocolumn,apj]{emulateapj} 
\usepackage{times}

\usepackage{lscape}

\newcommand{\eg}{{\it e.g.,}}
\newcommand{\ie}{{\it i.e.,}}
\newcommand{\kms}{{km s$^{-1}$}}
\newcommand{\cc}{{cm$^{-3}$}}
\newcommand{\othree}{{[O\,{\sc iii}]}}
\newcommand{\otwo}{{[O\,{\sc ii}]}}

\shorttitle{Extended Emission-Line Region of 4C 37.43} 
\shortauthors{Fu and Stockton}
\journalinfo{Accepted by the Astrophys. J.}
\submitted{Accepted by the Astrophys. J.}

\begin{document}
\title{
Integral Field Spectroscopy of the Extended Emission-Line Region of 4C
37.43\footnote{Based on observations obtained at the Gemini Observatory,
which is operated by the Association of Universities for Research in
Astronomy, Inc., under a cooperative agreement with the NSF on behalf of
the Gemini partnership: the National Science Foundation (United States),
the Particle Physics and Astronomy Research Council (United Kingdom),
the National Research Council (Canada), CONICYT (Chile), the Australian
Research Council (Australia), CNPq (Brazil) and CONICET (Argentina).
Some of the data presented herein were obtained at the W.M.
Keck Observatory, which is operated as a scientific partnership among
the California Institute of Technology, the University of California and
the National Aeronautics and Space Administration. The Observatory was
made possible by the financial support of the W.M. Keck Foundation.}  }
\author{Hai Fu and Alan Stockton\altaffilmark{2}}
\affil{Institute for Astronomy, University of Hawaii, 2680 Woodlawn
Drive, Honolulu, HI 96822}
\altaffiltext{2}{Also at Cerro Tololo Inter-American Observatory, Casilla 603, La Serena, Chile}

\begin{abstract}
 
We present Gemini integral field spectroscopy and Keck II longslit
spectroscopy of the extended emission-line region (EELR) around the
quasar 4C\,37.43. 
The velocity structure of the ionized gas is complex and cannot be
explained globally by a simple dynamical model. 
The spectra from the clouds are inconsistent with shock or ``shock +
precursor'' ionization models, but they are consistent with
photoionization by the quasar nucleus. 
The best-fit photoionization model requires a low-metallicity (12+log(O/H)
$\lesssim$ 8.7) two-phase medium, consisting of a matter-bounded diffuse
component with a unity filling-factor ($N \sim 1$ \cc, $T \sim 15000$
K), in which are embedded small, dense clouds ($N \sim 400$ \cc, $T \sim
10^4$ K). The high-density clouds are transient and can be re-generated
through compressing the diffuse medium by low-speed shocks ($V_S
\lesssim 100$ \kms). 
Our photoionization model gives a total mass for the ionized gas of
about $3\times10^{10}$ M$_{\odot}$, and the total kinetic energy implied by this
mass and the observed velocity field is $\sim2\times10^{58}$ ergs.  The
fact that luminous EELRs are confined to steep-spectrum radio-loud
QSOs, yet show no morphological correspondence to the radio jets,
suggests that the driving force producing the 4C\,37.43 EELR was a
roughly spherical blast wave initiated by the production of the jet. That
such a mechanism seems capable of ejecting a mass comparable
to that of the total interstellar medium of the Milky Way suggests
that ``quasar-mode'' feedback may indeed be an efficient means of
regulating star formation in the early universe.

\end{abstract}


\section{Introduction}

A number of low-redshift QSOs show luminous extended emission-line regions
(EELRs) that have characteristic dimensions of a few tens of kpc, considerably
larger than the typical $\sim1$ kpc of the classical narrow-line regions 
(NLRs; see \citealt{sto06b} for a recent brief review of EELRs).  In a few cases, 
such extended emission may be simply due to QSO photoionization of the 
{\it in situ} interstellar medium (ISM) of the host galaxy or of nearby gas-rich dwarf
galaxies, but, for the most luminous examples, such explanations are inadequate.
These luminous EELRs typically show complex filamentary structures that
bear no close morphological relationships either with the host galaxies or
with extended radio structures.

In spite of this general lack of attention on the part of EELRs to the structural
parameters of their host galaxies, they quite clearly are not oblivious to the
properties of the QSOs themselves.  As \citet{bor84} and \citet{bor85} found,
and \citet{sto87} confirmed, luminous EELRs are associated virtually exclusively
with steep-spectrum radio-loud QSOs.  \citet{sto87} also noted a strong
correlation between the strength of the nuclear narrow-line emission and
that of the extended emission.
This latter correlation may not be particularly surprising.  However, the restriction
of luminous EELRs to steep-spectrum quasars, together with the general lack
of correspondence between the distribution of the ionized gas and the radio
structure, presents an intriguing puzzle.  To add to the confusion, not all
low-redshift, steep-spectrum quasars (even among those that have luminous
classical NLRs) show luminous extended emission: the
fraction that do is $\sim$1/3 to 1/2 \citep{sto87}.

Early guesses regarding the origin of the extended gas centered on tidal
debris \citep{sto87} or cooling flows \citep{fab87}.  However, as \citet{cra88}
correctly pointed out, ionized gas from a tidal encounter would 
dissipate on a time scale of $\lesssim10^6$ years unless it were confined 
gravitationally or by the pressure of a surrounding medium.  On the other
hand, \citet{sto02} showed, from a detailed photoionization model of the
EELR around the $z = 0.37$ quasar 4C\,37.43, that the pressure of
any surrounding hot gas was too low for the gas to cool in less that a
Hubble time.  This conclusion was supported by the lack of evidence for
a general distribution of hot gas around 3 quasars in deep {\it Chandra}
X-ray imaging \citep{sto06}.
Gravitational confinement of the amount of ionized gas seen in many
EELRs would imply unreasonably large masses of neutral or otherwise
invisible material in the outskirts of the host galaxy.

Given the short lifetime of unconfined ionized filaments and the unlikelihood
of gravitational confinement, along with the other constraints we have 
mentioned, the most likely scenario for producing an EELR
is that of a superwind
\citep{sto02,fu06}.  In principle, such a superwind could result either from
a starburst or from feedback from the quasar (\eg\ \citealt{diM05,hop06}), 
with some evidence favoring the latter alternative \citep{fu06}.  Such
an origin would be of considerable interest, since it would mean that
we have local examples of a process similar (though almost certainly
not identical) to the ``quasar-mode''
feedback mechanism that may be important in initially establishing the observed
correlation between bulge mass and black hole mass during galaxy formation
in the early universe.  In order to explore this possibility in more
detail, we have carried out an extensive re-examination of the EELR of
4C\,37.43.

\begin{figure*}[!tb]
\epsscale{1.0}
\plotone{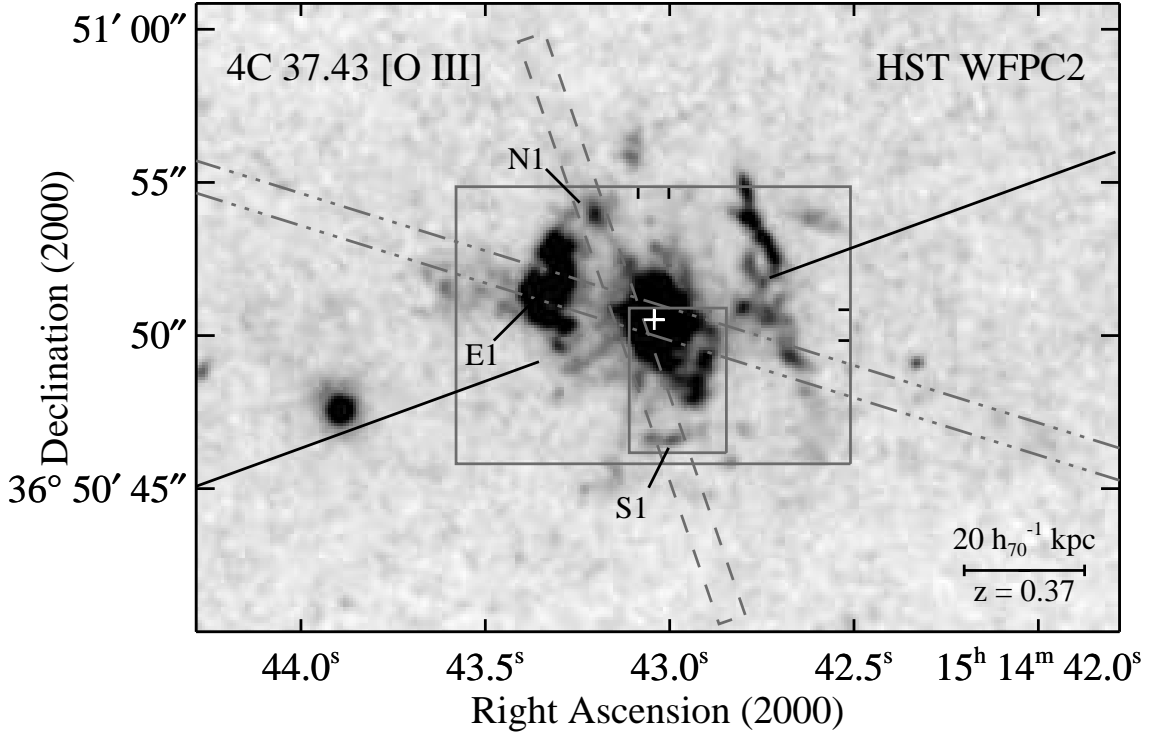}
\caption{
{\it Hubble Space Telescope} {\it (HST)} WFPC2 \othree\ image of 4C\,37.43 \citep{sto02}.  Overlaid are
the field-of-view (FOV) of the mosaicked GMOS IFU2 datacube (outer solid box), the field of
the GMOS IFUR datacube (inner solid box), the position of the DEIMOS
slit (long dashed rectangle) and that of the LRIS slit (long dash dotted
lines).  
The tick marks on the inner edge of the total IFU2 FOV show the overlapping
regions of the four IFU2 pointings; from each corner, the rectangle defined
by the farthest tick mark from the corner gives the FOV for a single
pointing.
The radio jet direction is described by the solid lines.  The
cross near the center marks the position of the QSO.  
} \label{fig:obs} \end{figure*} 

4C\,37.43 has the most luminous EELR among QSOs at $z
\leq 0.45$ \citep{sto87}. The continuum image of the host galaxy shows
distortions and a likely tidal tail that indicate that a major merger is in
progress, which has probably triggered the current episode of quasar
activity. Like most others, this EELR is morphologically
completely uncorrelated with both the stellar distribution in the host galaxy and with
the FR II double radio source \citep{mil93}. It has two main
condensations of ionized gas at roughly the same distance
($\sim$3.5\arcsec, \ie\ about 18 kpc) to the east and west of the quasar
(Fig.~\ref{fig:obs}). The optical spectrum of the brightest region, the
main east condensation (E1; following \citealt{sto02}), has
been discussed in several papers before \citep{sto76,bor84,sto02}. Here
we present our new GMOS integral field spectroscopy and DEIMOS longslit
spectroscopy covering a large fraction of the whole EELR. 
Throughout we assume a flat cosmological model with $H_0=70$ km
s$^{-1}$ Mpc$^{-1}$, $\Omega_m=0.3$, and $\Omega_{\Lambda}= 0.7$.  

\section{Observations and data reduction} 
\subsection{GMOS Integral Field Spectroscopy} \label{sec:obs_gmos}

4C\,37.43 was observed with the Integral Field Unit (IFU;
\citealt{all02}) of the Gemini Multiobject Spectrograph (GMOS;
\citealt{hoo04}) on the Gemini North telescope. 
We observed a $13\arcsec \times 9\arcsec$ region centered on the QSO in
the early half-night of 2006 May 23 (UT). Since the main purpose of
these observations was to determine the global velocity field of the
EELR through the \othree\,$\lambda$5007 line, the IFU was used in the
full-field (two-slit) mode and dithered on a rectangular grid of
$6\arcsec \times 4\arcsec$. 
The QSO was successively placed at each corner of the $7\arcsec\times5\arcsec$
IFU field and $\sim$0\farcs5 away from the edges, so
the fields covered by the pointings overlap with each other by
$\sim$1\arcsec\ (see Fig.~\ref{fig:obs}).
We obtained two 720-s frames and one 156-s frame on the SE
pointing\footnote{The third exposure was cut short due to a dome
problem.} and three 720-s exposures on the other three pointings.
With the R831/G5302 grating and a central wavelength of
658.5 nm, we obtained a dispersion of 0.34\AA\ per pixel, a spectral
resolution of $\sim1.3$\AA\ (58 \kms) FWHM, and a wavelength range of
6330--6920 \AA.  The $r$ and RG610 filters were used to avoid spectral
overlaps. The spectrophotometric standard star Feige 34 was observed for
flux calibration. The seeing was $\sim0\farcs4$ throughout the
half-night. 
Before running the data through the reduction pipeline (see the
3rd paragraph in this section), the exposures for each pointing position were
combined using the IRAF task IMCOMBINE. We weighed the frames according
to their integration times. Pixels were rejected if their values were
7-$\sigma$ off the median\footnote{$\sigma$ estimated from known CCD
parameters.}. For all four positions, there are no apparent cosmic rays
in the final image. The shortened exposure of the SE pointing did make
the data a little bit shallower in this region than others. We corrected
for this difference by using an exposure map while making the final
mosaicked datacube. 

We also obtained deep integral field spectroscopy on a $\sim3\farcs5
\times 5\arcsec$ region about 2\arcsec\ SW of the QSO. A total of five
2400-s exposures were taken with the half-field (one-slit) mode in the
first half-night of 2006 May 24 (UT). This configuration provides a
wider spectral coverage but a smaller field than does the two-slit mode. 
With the B600/G5303 grating and a central wavelength
of 641.2 nm, the dispersion and spectral resolution were approximately
0.46\AA\ per pixel and 1.8\AA\ (FWHM). The wavelength range was 4250--7090 
\AA, which includes emission lines from [Ne\,{\sc
v}]\,$\lambda$3426 to \othree\,$\lambda$5007. BD\,+28\arcdeg\,4211 was observed
for flux calibration. The seeing was $\sim$0\farcs6 during the
half-night. 

The data were reduced using the Gemini IRAF package (Version 1.8). The
data reduction pipeline (GFREDUCE) consists of the following standard
steps: bias subtraction, cosmic ray rejection, spectral extraction,
flat-fielding, wavelength calibration, sky subtraction, and flux
calibration. Spectra from different exposures were assembled and
resampled to construct individual datacubes (x,y,$\lambda$) with a pixel
size of 0\farcs05 (GFCUBE). For each datacube, differential atmosphere
refraction was corrected by shifting the image slices at each wavelength
to keep the centroid of the quasar constant. The four datacubes resulting
from the four IFU two-slit pointings were then merged to form a single
datacube (the IFU2 datacube), and the five datacubes from the IFU
one-slit mode were combined to form the IFUR datacube. Finally, these
two datacubes were binned to 0\farcs2 pixels, which is the original
spatial sampling of the IFU fiber-lenslet system. 

Since our study focuses on the emission-line gas, it is desirable to
remove the light of the quasar from the datacubes. Since the
emission-line clouds show essentially pure emission-line spectra, we
used the continuum on either side of an emission line to precisely
define the PSF of the quasar. We then ran the two-component
deconvolution task PLUCY (contributed task in IRAF; \citealt{hoo94}) on
each image slice across the emission line to determine the flux of the
quasar at each wavelength. Finally the quasar component was removed by
subtracting a scaled PSF (according to the fluxes determined by PLUCY)
from each image slice. Note that the deconvolved images produced by
PLUCY were not used to form the QSO-removed datacubes. 
       
\subsection{DEIMOS Longslit Spectroscopy} \label{sec:obs_deimos}

We obtained additional longslit spectroscopy using the DEep Imaging
Multi-Object Spectrograph (DEIMOS; \citealt{fab03}) of the Keck II
telescope on 2006 August 28 (UT). 
We initially centered the 20\arcsec\ long slit on the QSO at a position
angle of 19$^{\circ}$ (N to E), then offset it by 0\farcs73 so that it
goes through two EELR clouds (N1 \& S1; Fig.~\ref{fig:obs}) that are
presumably associated with X-ray emission (see Fig.~5 in
\citealt{sto06}). The alignment accuracy of DEIMOS is typically
$\sim$0\farcs1\footnote{http://www2.keck.hawaii.edu/inst/deimos/specs.html}. 
The total integration time was 3600 s. With a 600
groove mm$^{-1}$ grating, a $0\farcs9$ wide slit and a central
wavelength at 7020\AA, we obtained a dispersion of 0.63 \AA\ per pixel,
a resolution of 3.5 \AA, and a spectral range of 4570--9590 \AA, covering
emission lines from [Ne\,{\sc v}]\,$\lambda$3346 to [S\,{\sc
ii}]\,$\lambda$6731. The spectrophotometric standard star Wolf\,1346 was
observed with a 1\farcs5 wide slit at parallactic angle. The GG400
filter was used for all observations. The seeing was $\sim0\farcs7$
during the observations.
We reduced the data using the $spec2d$ data reduction 
pipeline\footnote{The $spec2d$ pipeline was developed at UC Berkeley with
support from NSF grant AST-0071048.}. The standard star spectrum suffers
second-order contamination in the spectral range $\gtrsim$ 8000 \AA;
for the spectra of the EELR this contamination is negligible since
there is essentially no continuum. However, in order to flux calibrate the
emission-line spectra, one needs to correct for the second-order
contamination of the standard star spectrum. We thus obtained a DEIMOS
spectrum of Feige\,34 taken on November 29 2006 (UT). This standard star
was observed using the same settings as ours except that a different
filter (GG495) and a slightly different central wavelength were used.
Since both GG400 and GG495 filters reach an identical transmission
value redward of 7000 \AA, no correction for the transmission curves
needs to be applied. We first derived the calibration files with
STANDARD for both spectra, then
multiplied the Feige 34 data by a constant 
so that the two matched between 7000 and 7600 \AA. Then, in the bluer region,
we selected the data points from the GG400 spectrum (Wolf\,1346), while in
the redder region, we selected the GG495 data (Feige\,34). We then fitted a smooth
curve to the selected data to form the final sensitivity
function (SENSFUNC), which was used in the flux calibration. 

Proper sky subtraction is often a problem beyond 7000 \AA\ because of
strong airglow lines and fringing in the CCD. The $spec2d$
pipeline attempts to defringe the spectra by dividing the science spectra
with a normalized flat field. This feature significantly improves the
sky subtraction result, although it is still not perfect when dealing
with the strongest sky lines. Fortunately, the redshift of 4C\,37.43 nicely places the
important red lines (\ie\ [N\,{\sc ii}]\,$\lambda\lambda$6548,6584,
H$\alpha$ and [S\,{\sc ii}]\,$\lambda\lambda$6717,6731) in a window (8900--9300
\AA) between two strong OH bands that is essentially free
of strong sky lines, making the subtraction of sky lines almost perfect
in this region.

Figure~\ref{fig:obs} presents an overview of the regions covered by
the observations discussed above. We also include a deep Keck I LRIS
(Low-Resolution Imaging Spectrometer; \citealt{oke95}) spectrum from
\citet{sto02} in our discussion. That spectrum had a total integration of
3600 s and was taken with a 300 groove mm$^{-1}$ grating and a
1\arcsec\ wide slit.  The position of this slit is also shown in
Fig.~\ref{fig:obs}.

\section{Results}

\subsection{Kinematics} \label{sec:kinematics}

\begin{figure*}[!tb]
\epsscale{1.0}
\plotone{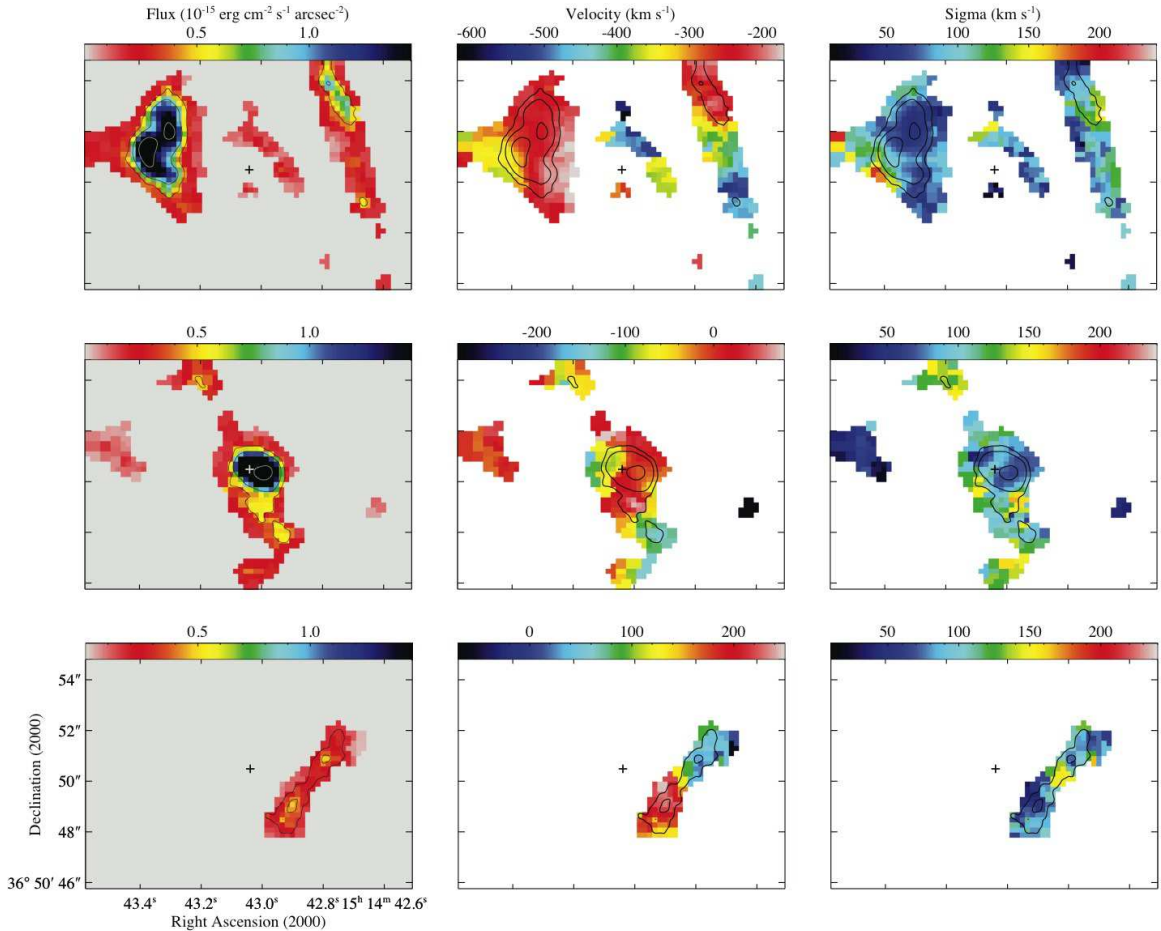}
\caption{
Velocity field of 4C\,37.43 EELR derived from the \othree\,$\lambda$5007
region of the QSO-subtracted IFU2 datacube.  The three columns, from left to
right, are line intensity, radial velocity (relative to that of the
nuclear NLR) and velocity dispersion maps. To separate
different clouds that are present along the same line of sight, the
velocity field is shown in three separate velocity bands ($-$620 to
$-$170 \kms, $-$290 to +80 \kms\ and $-$70 to +250 \kms\ from top to
bottom), and the line intensity and velocity dispersion maps refer to the
same clouds. Pixels are 0\farcs2 squares.  The crosses indicate the position
of the quasar before it was removed from the datacube (for details see
\S~\ref{sec:obs_gmos}). Contours are from the line intensity maps.
} \label{fig:vfield} \end{figure*} 

Figure~\ref{fig:vfield} presents the velocity field of the
\othree\,$\lambda$5007 emission derived from the IFU2 datacube, which
has a spectral range that brackets the H$\beta$--\othree\ region. The
QSO was removed from the datacube using the method described in
\S~\ref{sec:obs_gmos}. We discarded spectra that show very low
amplitude-to-noise ratios (A/N $\leq$ 3), and spatially binned the
remaining spectra with A/N $\leq$ 10 to a target A/N $\approx$ 8 using
the Voronoi binning method of \citet{cap03}. We then used multiple
Gaussians to fit the \othree\,$\lambda$5007 line profile in the binned
datacube.  Because different velocity components are sometimes present
along the same line of sight, the velocity field is displayed in three
velocity bands from $-$620 to +250 \kms (relative to $z_0 = 0.37120$,
the redshift of the quasar NLR as determined from its
\othree\,$\lambda\lambda$4959,5007 lines). The velocity dispersion
($\sigma$) measurements were corrected for the $\sigma_0=24.6$ \kms\
instrumental resolution. 
Monte-Carlo simulations were performed to determine the uncertainties in
the measured line parameters. For A/N = 8 and an intrinsic $\sigma = 50$
\kms\ (130 \kms), the 1-$\sigma$ errors on velocities ($V$) and velocity
dispersions ($\sigma$) are both about 4 \kms (6 \kms), and the errors on
fluxes are about 6\% (4\%).

Our velocity maps are generally consistent with those from previous
integral-field spectroscopy and image-sliced spectroscopy
\citep{dur94,cra00,sto02}, but they offer a much clearer view of the
complexities in the velocity structure of this region. 
Overall, these maps show that: 
(1) the clouds comprising the EELR are certainly not in a coherent
rotation about the QSO, as already pointed out by \citet{sto76} with
very limited velocity information; 
(2) the majority of the clouds are blueshifted relative to the QSO, and
the highest blueshifted velocity is about $-$620 \kms, while the highest
redshifted velocity is only +250 \kms;
(3) the velocity dispersions for the most part are between 50 and 130
\kms (or 120 to 310 \kms\ FWHM), much higher than the sound speed in a
hydrogen plasma at $10^4$ K, $c_s \sim$ 17 \kms;
(4) there is no obvious evidence for jet-cloud interactions. As seen in
Fig.~\ref{fig:obs}, the radio jet just misses the two major
condensations, and no significant increase in velocity dispersion is
observed where the jet crosses any visible part of a cloud.

As shown in the $-620$ to $-170$ \kms\ panels, the two major
condensations to the east (E1) and west of the QSO are both blueshifted
to about $-250$ \kms. E1 is resolved in our velocity maps, thanks to the
0\farcs2 sampling of the IFU and the 0\farcs4 seeing. The southern half
of E1 shows a velocity increase from $-$180 to $-$350 \kms\ along the W-E
direction, and also an increase in $\sigma$ from 50 to 110 \kms\ along
the same direction. This kinetic structure suggests that E1 is not a coherent body (first suspected by \citealt{dur94}); instead, it is composed of sub-clouds, as clearly seen in the {\it HST} \othree\ image (see the low-contrast insets to Fig.~3 of \citealt{sto02}).

Overall, the velocity field of the 4C\,37.43 EELR is similar to that of the
EELR around 3C\,249.1 \citep{fu06}, in the sense that both of them
appear globally chaotic but locally ordered. 

\subsection{Electron Density and Temperature} \label{sec:ne_t}

\begin{figure*}[!tb]
\epsscale{1.0}
\plotone{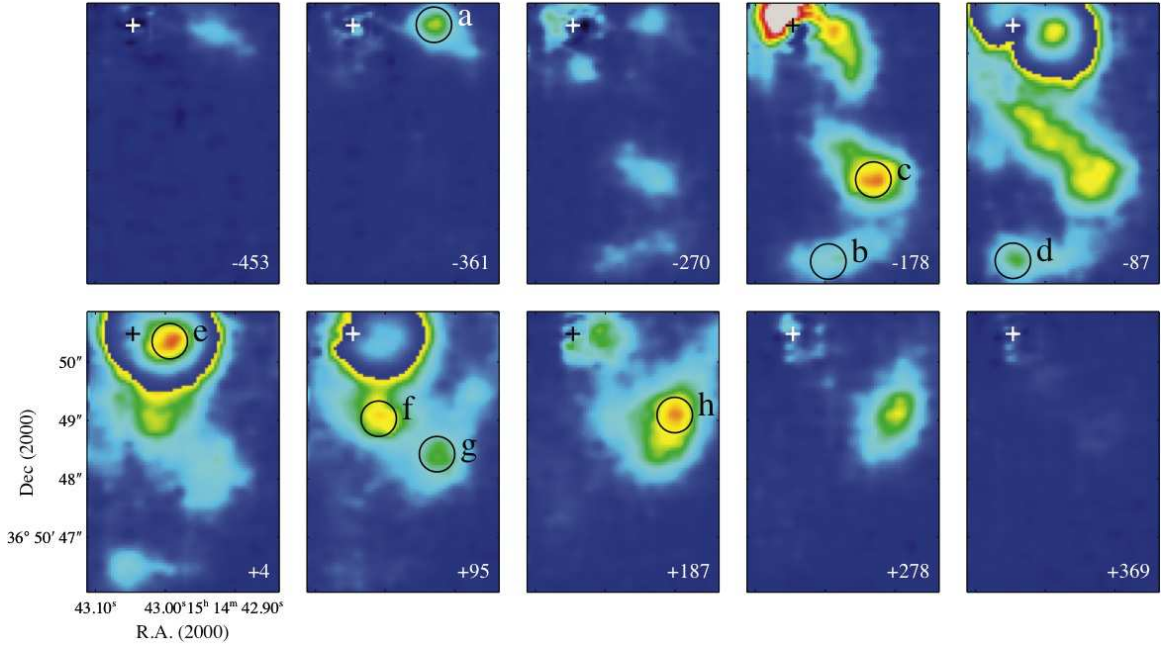}
\caption{
\othree\,$\lambda$5007 radial velocity channel maps from the QSO-subtracted
IFUR datacube.  The central velocities, in \kms\ relative to that of the
quasar NLR, are indicated in the lower-right corner of each panel. The circles show the extraction apertures of the various emission-line
clouds discussed in the text. Some of the images have been allowed to
wrap around to show both low surface brightness detail and high surface
brightness peaks.  The crosses indicate the position of the QSO before
it was removed from the datacube.  
} \label{fig:exreg} \end{figure*} 

Our IFUR datacube covers the \otwo\,$\lambda\lambda$3726,3729 doublet, and
its spectral resolution ($\sim$1.8\AA) is high enough to resolve the
two. We can determine the \otwo\ luminosity-weighted average electron
densities ($N_e$\otwo) from the intensity ratio of the doublet
\citep{ost89}. Electron temperatures ($T_e$) can be measured using the
\othree\,($\lambda$4959+$\lambda$5007)/$\lambda$4363 ratio (R$_{\rm OIII}$).  
Most of the
clouds in the IFUR field-of-view (FOV) have a low surface brightness, making it
impossible to reliably measure the line ratios from a single 0\farcs2
pixel. We therefore identified the peaks of the individual clouds, then binned
pixels within $0\farcs3$ of the peak. Figure~\ref{fig:exreg} displays
the extraction apertures for the eight resolved clouds.

\begin{figure*}[!tb]
\epsscale{1.0}
\plotone{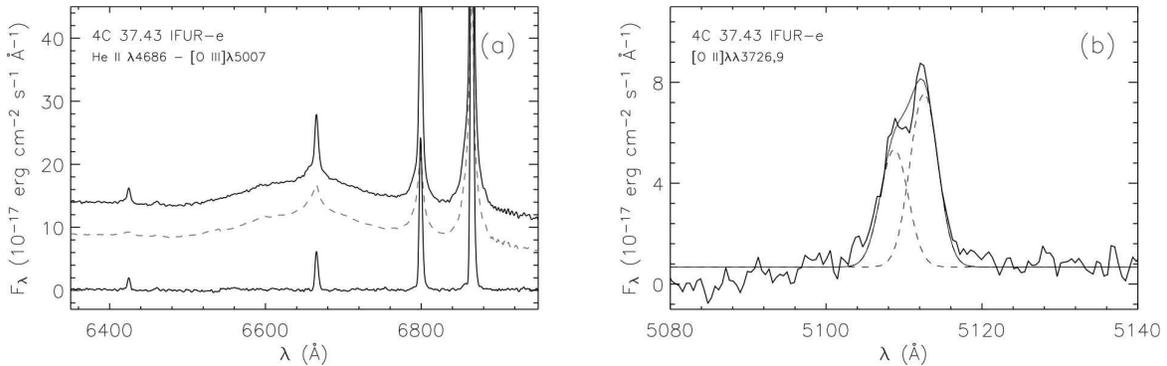}
\caption{
Spectra of 4C 37.43 EELR {\it e} extracted from the IFUR datacube.
({\it a}) The He\,{\sc ii}\,$\lambda4686$ to \othree\,$\lambda5007$
portion of the spectra. The upper and lower solid curves are the
extracted spectra before and after removing the QSO from the 
datacube. The dashed curve shows the difference between the two, \ie\
the scattered light from the QSO. For clarity, the dashed curve was
shifted downward by $5\times10^{-17}$ erg cm$^{-2}$ s$^{-1}$ \AA$^{-1}$.
({\it b}) The \otwo\,$\lambda\lambda$3726,9 doublet of the EELR {\it
e}. QSO scattered light has been removed. The dashed curves show the
decomposition of the doublet, and the smooth solid curve is the sum of
the two.
} \label{fig:spec} \end{figure*} 

Figure~\ref{fig:spec}{\it a} compares the extracted spectra from cloud
$e$ before and after removing the QSO from the datacube
(\S~\ref{sec:obs_gmos}). It shows that the PSF subtraction technique can
successfully remove the QSO scattered light even for a cloud only
0\farcs6 from the quasar, as confirmed by the absence of the broad
H$\beta$ line in the residual spectrum. Note that since the data were
taken under excellent seeing condition ($\sim 0\farcs6$), PSF subtraction
is only critical for the clouds very close to the quasar, namely $a$,
$e$, and $f$. In addition, the line ratios for clouds $a$ and $f$ should
be reliable even if the QSO removal were just barely correct, because
the two are well separated from the quasar NLR in
velocity. We are less confident about the line ratios of cloud $e$,
since its velocity is almost centered on that of the NLR. However, the
shape of the cloud as seen in the residual datacube (see
Fig.~\ref{fig:exreg} as an example) suggests that the PSF removal cannot
be too far off. Also at least the \otwo\ ratio should be trustworthy
since the NLR emits very weakly in this doublet.

To estimate the effect of using a fixed-size aperture on the covering
fraction of the PSF (since the spatial resolution gets better as the
wavelength increases), we compared a QSO spectrum extracted from a
$d=4\farcs8$ aperture with another one from a $d=0\farcs6$
aperture\footnote{Because the QSO is positioned at the upper left corner
of the IFUR FOV, the extracted spectra were from the lower right quarter
of the QSO only. We of course had to assume that the PSF was
symmetric.}. A smoothed version of the ratio of the two was used to
correct for the spectra of all the emission-line clouds. 
The atmospheric B-band absorption was corrected by dividing the spectra
by an empirically determined absorption law. 
The Galactic extinction on the line of sight to 4C\,37.43 is $A_V$ =
0.072 \citep{sch98}. 
Intrinsic reddening due to dust associated with the cloud, whenever
possible, was determined from the measured H$\gamma$/H$\beta$ ratio
assuming an intrinsic ratio of 0.468 (for case B recombination at T =
$10^4$ K and $N_e = 10^2$ \cc; \citealt{ost89}).  However, there are
four clouds, namely $a, b, d,$ and $g$, where H$\gamma$ is not securely
detected in the IFU data. For clouds $b$ and $d$, we used the reddening implied by the
H$\gamma$/H$\beta$ ratio from the DEIMOS spectrum of S1
(Fig.~\ref{fig:obs}). For $a$ and $g$, we assumed an arbitrary reddening
of $A_V=0.7$.  Both reddening effects were corrected using a standard
Galactic reddening law \citep{car89}. 

To measure the line ratios of the \otwo\ doublet, we used two Gaussian
profiles constrained to be centered on the expected wavelengths of the
doublet after applying the same redshift.  As an example,
Fig.~\ref{fig:spec}{\it b} shows the profile of the \otwo\ doublet for
the cloud $e$ and the best two-Gaussian model fit.  For the cases where
there are two velocity components in a single spectrum, we first
identify which component corresponds to the target cloud and which one is
from contaminating light from other nearby clouds. Then we measure the
velocity difference between the two from the
\othree\,$\lambda\lambda$4959,5007 lines. Finally we freeze this parameter
when fitting other lines, including the \otwo\ doublet. For the DEIMOS
spectra of N1 and S1, the \otwo\ doublet is unresolved, so we measured
the density-sensitive [S\,{\sc ii}]\,$\lambda$6716/$\lambda$6731 ratios. 

When there is a good \othree\,$\lambda$4363 line measurement, we derived
$T_e$ and $N_e$ consistently using the IRAF routine TEMDEN. Otherwise, a
uniform $T_e = 10^4$ K was assumed. $N_e$\otwo ($N_e$[S\,{\sc ii}])
would be approximately 15\% (5\%) higher, if $T_e$ = 15000 K.

The physical properties of the EELR clouds are summarized in
Table~\ref{tab:prop}. The listed \othree\,$\lambda$5007 velocity
dispersions have been corrected for the instrumental broadening
($\sigma_0 \approx 35$ \kms\ for IFUR; $\sigma_0 \approx 67$ \kms\ for
DEIMOS). Reddening-free emission-line fluxes and 3-$\sigma$ upperlimits
are tabulated in Table~\ref{tab:liner} as ratios to the H$\beta$ flux. 
The quoted 1-$\sigma$ errors were derived from the covariance matrices
associated with the Gaussian model fits, with the noise level estimated
from line-free regions on either side of an emission line.
For completeness, we have also included E1 in both tables. Most of the line
fluxes were re-measured from the LRIS spectrum except for the \otwo\
doublet ratio, which is quoted from \citet{sto02}. 

In general, the temperatures of the clouds are about
15000 K, and the densities are a few hundreds \cc. The cloud $e$ looks
peculiar in this crowd: it has both the coolest temperature and the
lowest density, implying a pressure much lower than the
average. The only other low-density cloud is N1, where the [S\,{\sc ii}]
ratio is at the low-density limit. 

\subsection{Ionization Mechanisms} \label{sec:ionization}

\begin{figure*}[!tb]
\epsscale{1.0}
\plotone{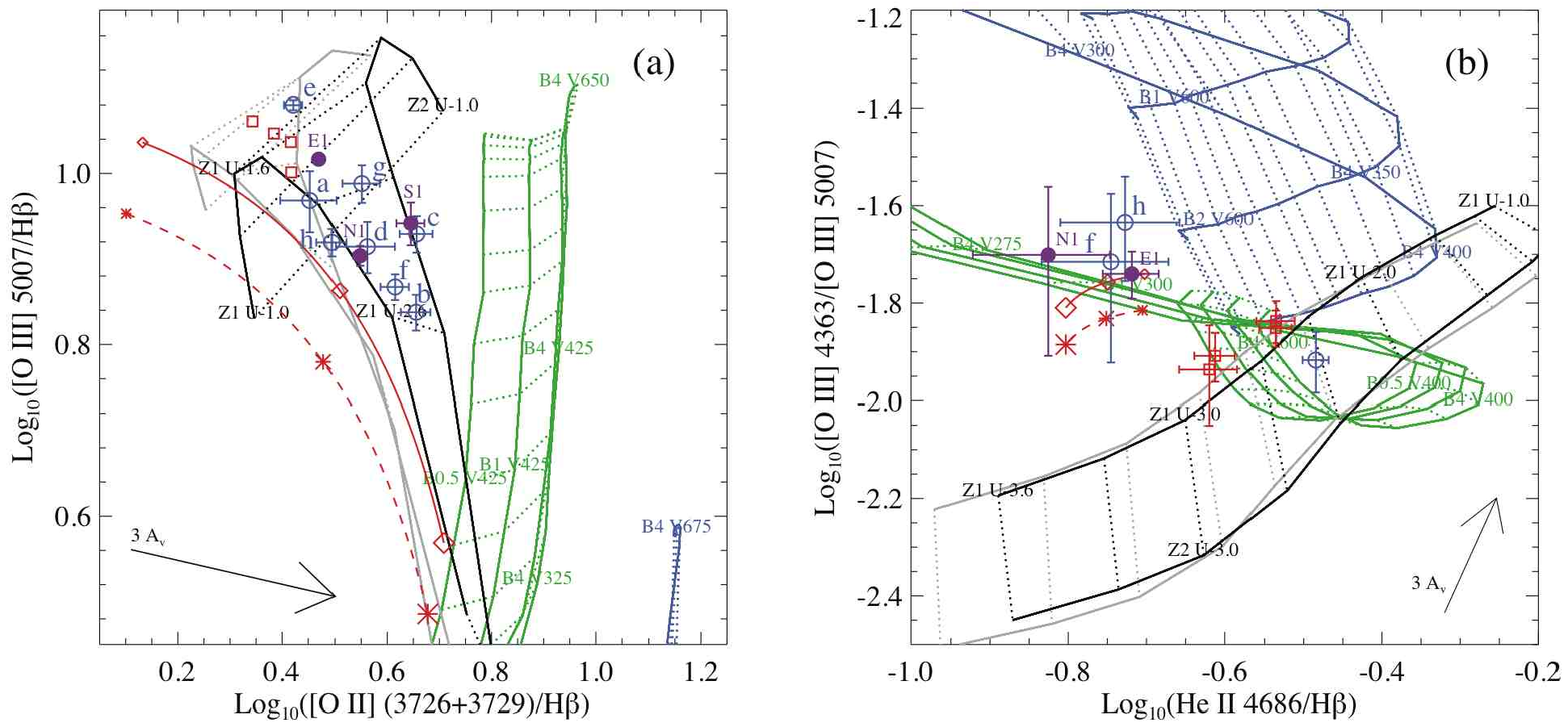}
\caption{
Line ratio diagnostic diagrams distinguishing different ionization
mechanisms: (a) \otwo\,$\lambda\lambda3726,9$/H$\beta$\ 
vs. \othree\,$\lambda5007$/H$\beta$, 
(b) He\,{\sc ii}\,$\lambda4686$/H$\beta$ vs.
\othree\,$\lambda4363$/\othree\,$\lambda5007$, and 
(c) \otwo\,$\lambda\lambda3726,9$/[Ne\,{\sc iii}]\,$\lambda3869$\ vs.
\othree\,$\lambda5007$/H$\beta$.
Overplotted are model grids from the dusty radiation-pressure dominated
photoionization model (black and grey; \citealt{gro04}), shock-only
(blue) and ``shock + precursor" models (green; \citealt{dop96}), along
with the two-component photoionization sequences with two different abundance sets
(red lines with symbols).
The dusty photoionization grids cover a range of ionization parameters
(log($U$)) for metallicities of 1.0 and 2.0 $Z_\odot$, and two
power-law indexes for the ionizing continuum (black grids for $\alpha =
-1.2$, and grey ones for $\alpha = -1.4$). 
In the shock models, a range of shock velocities ($V_S$ [km s$^{-1}$])
and magnetic parameters ($B/n^{1/2}$ [$\mu$ G cm$^{-3/2}$] ) are
covered.
The two-component photoionization sequences show a range of combination
ratios between the two components. The dashed and solid red lines
represent the models adopting gas abundance set $a$ and set $b$,
respectively (see \S~\ref{sec:ionization} for details).  The asterisks
and diamonds, with increasing sizes, denote the positions where 20, 50
and 80\% of the total H$\beta$ flux is from the high-density IB
component.  
Line ratio measurements from the 4C\,37.43 IFUR datacube are shown as
blue open circles, and purple filled circles show the measurements from
either LRIS (E1) or DEIMOS (N1 \& S1) longslit spectra. For comparison,
the four EELR clouds of 3C\,249.1 \citep{fu06} are plotted as red open
squares. 
Arrows are reddening vectors.  
} \label{fig:lratio} \end{figure*} 

\addtocounter{figure}{-1}
\begin{figure}[!tb]
\epsscale{1.0}
\plotone{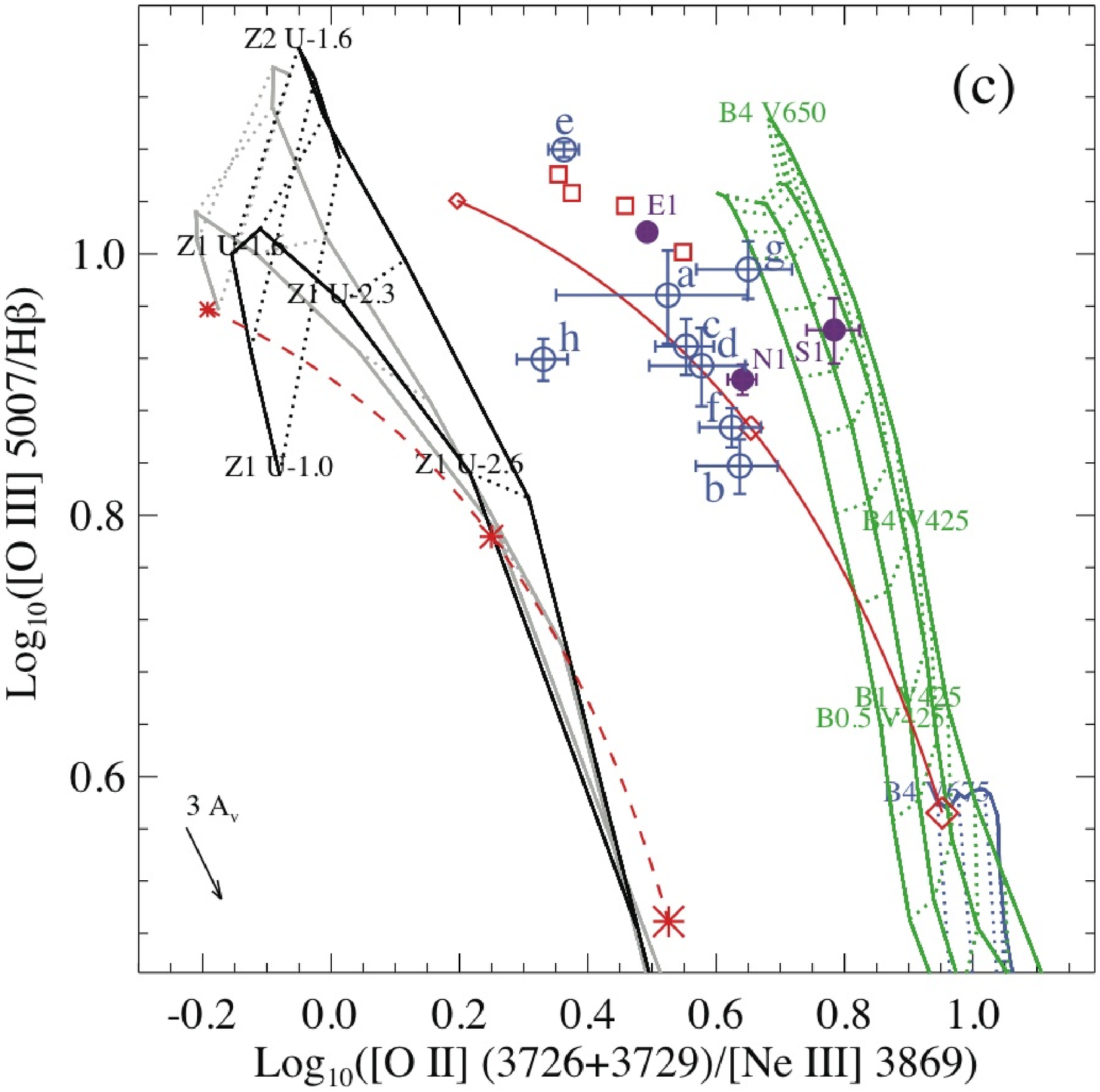}
\caption{
{\it continued}
} \end{figure}

Line ratio diagnostic diagrams are widely used to distinguish different
ionization mechanisms and to infer useful physical parameters of ionized
clouds.  Figure~\ref{fig:lratio} presents three of these diagrams.  Our
line-ratio data are plotted against four ionization models, which were
all computed by the multipurpose photoionization-shock code MAPPINGS: 
(1) a dusty radiation-pressure dominated photoionization model
\citep{gro04},
(2) a two-phase photoionization model,
(3) a shock ionization model and (4) a ``shock + precursor" model
\citep{dop96}. 

These diagrams clearly show that both pure shock and ``shock +
precursor" are not as successful as the photoinozation models in
reproducing the observed spectra of the EELR. At first sight,
Fig.~\ref{fig:lratio}{\it a} might seem to imply that the ``shock +
precursor" model would possibly fit the data if we weight the precursor
component more heavily. However, the high shock velocities ($>$ 400
\kms) required to reach the observed high \othree/H$\beta$ ratios are
inconsistent with the measured low velocity dispersions of those clouds
(Fig.~\ref{fig:vfield}). In addition, making the precursor more luminous
would only move the grids below where they are in
Fig.~\ref{fig:lratio}{\it b}, hence further off the data points.
Therefore, we believe that it is unlikely that any of these clouds are
dominated by shock ionization; instead, they are photoionized by the
central QSO.  In fact, it was argued before that at least E1 was
photoionized by the QSO, as indicated by the low intensity ratios of
O\,{\sc iv}\,$\lambda1402$/He\,{\sc ii}\,$\lambda1640$ and C\,{\sc
iv}\,$\lambda1549$/He\,{\sc ii} \citep{sto02}.

Briefly, Model 1 simulates the photoionization of the surface layer
($\lesssim$ a few pc) of a dense self-gravitating molecular cloud, where
the hydrogen density near the ionization front is about 1000 \cc\ and
the density structure is predominantly determined by radiation pressure
forces exerted by the absorption of photons by gas and dust. Detailed
discussions of this model can be found in \citet{gro04}.

The two-phase photoionization model is similar to the $A_{M/I}$ model by
\citet{bin96}. It assumes that the observed emission-line spectrum is a
combination of the emission from two species of clouds.
Guided by the results of \citet{sto02}, we used the latest version of
MAPPINGS (3r) to model two isobaric components: (1) a high-excitation
matter-bounded (MB) component with a density $N_H \sim 1$ \cc, a
temperature $T \sim 1.5\times10^4$ K and an average ionization parameter
$U \sim 6\times10^{-2}$, and (2) a low-excitation ionization-bounded
(IB) component with $N_H \sim 400$ \cc, $T \sim 10^4$ K and $U \sim
2\times10^{-4}$. Unlike the $A_{M/I}$ model, shielding of the ionizing
source by the MB component is not considered, \ie\, both clouds are
facing the same ionization field.
Different combination ratios of the two components produce a
sequence of spectra. We show in the diagrams a range of models having
20 to 80\% of the total H$\beta$ flux from the high-density IB
component.  

Both photoionization models used simple power laws to represent the
ionizing continuum ($F_{\nu} \propto \nu^{\alpha}$). 
Model 1 assumed a single-index power-law from 5 to 1000 eV, and we show
here two index values \citet{gro04} modelled, $\alpha = -1.2, -1.4$.  Model 2
uses a more realistic segmented power-law with $\alpha = -1.5$ in the UV
(8--185 eV) and $\alpha = -1.2$ in the X-ray (185--5000 eV). The
ionizing flux at the Lyman limit is normalized to $3.3 \times 10^{-17}$
erg cm$^{-2}$ s$^{-1}$ Hz$^{-1}$ \citep{sto02}.  These indexes are close
to the canonical value proposed for the typical spectral-energy
distribution of a QSO (\ie\ $\alpha = -1.4$; \eg\ \citealt{fer86}).

We have run our two-phase model on two different sets of abundances: (a)
the same dust-depleted 1 $Z_\odot$ abundance set as used in Model 1
\citep{gro04}; (b) the ``standard'' Solar abundance set ($Z^{'}_\odot$)
of \citet{and89} scaled by $-0.5$ dex, following \citet{sto02}. Note
that the total gas metallicities in the two sets are both around
one-third of $Z^{'}_\odot$. 
Table~\ref{tab:abund} compares the two abundance sets.
Sequences from both runs are shown in
Fig.~\ref{fig:lratio}. In all diagrams, set $b$ clearly provides a
superior fit to the data than set $a$, meaning that the gas abundance
in those clouds is more similar to $Z^{'}_\odot$. It also explains why
there is a drastic offset\footnote{This offset may imply that the Solar Ne/O abundance used by \citet{gro04} (log(Ne/O) = 0.61) is too high, by about a factor of three. In fact, new Solar wind measurements \citep{glo07} have updated the Solar Ne/O with a new value (log(Ne/O) = 1.12), which is almost exactly three times lower.}
between the data and the grids from Model 1 in
Fig.~\ref{fig:lratio}{\it c}. Therefore, hereafter the two-phase model (or
Model 2) only refers to the run using abundance set $b$.

\begin{figure*}[!t]
\epsscale{1.0}
\plotone{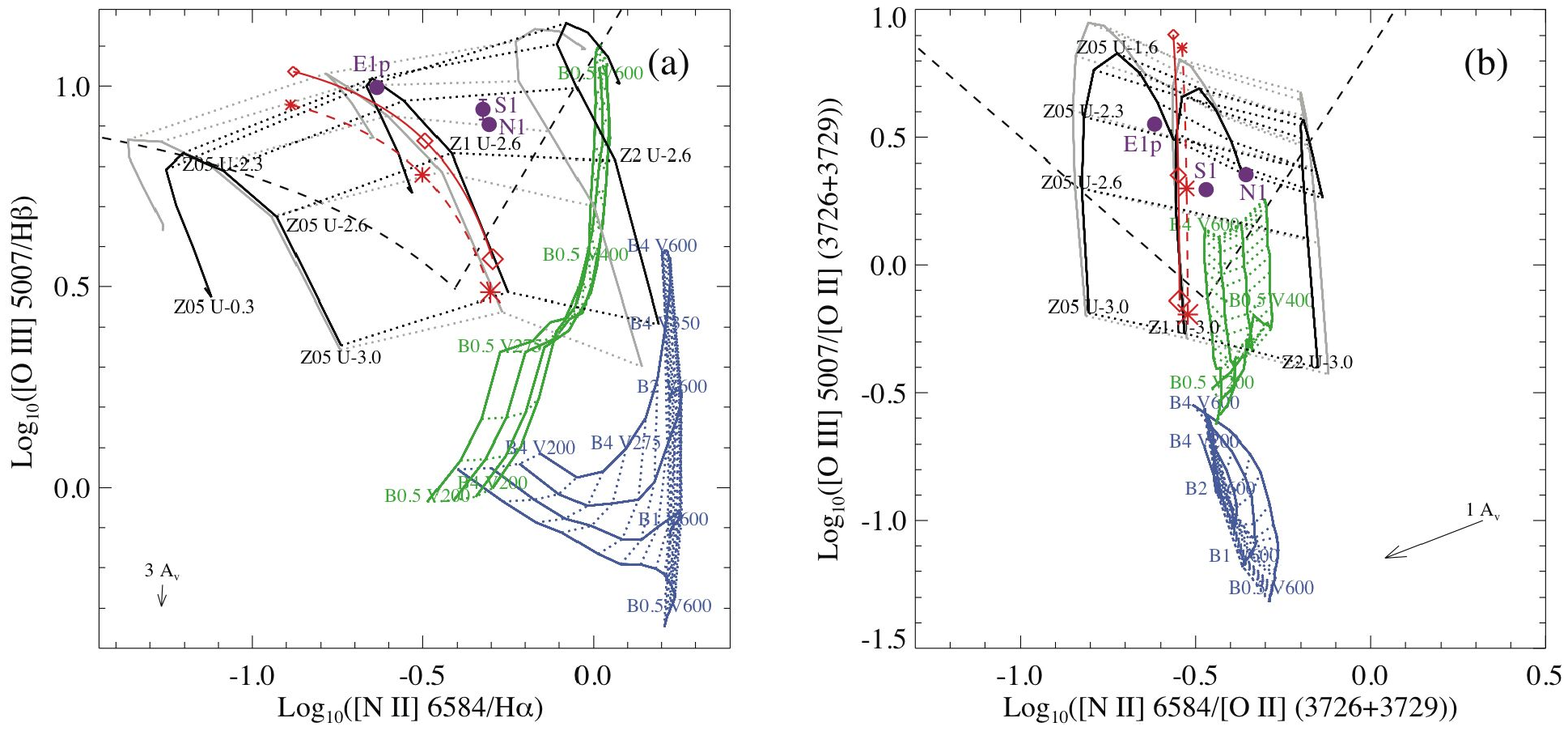}
\caption{
Metallicity sensitive diagnostic diagrams --- (a) [N\,{\sc
ii}]\,$\lambda6584$/H$\alpha$ vs. \othree\,$\lambda5007$/H$\beta$, 
(b) [N\,{\sc ii}]\,$\lambda6584$/\otwo\,$\lambda\lambda3726,9$ vs.
\othree\,$\lambda5007$/\otwo\,$\lambda\lambda3726,9$.
Keys are the same as in Fig.~\ref{fig:lratio}.  The photoionization
girds here display a smaller range of ionization parameters but a wider
range of metallicities (0.5, 1.0 and 2.0 $Z_\odot$).
In panel {\it a} the dashed black curve is the empirical starbursts/AGN
dividing line \citep{kew02}, and in panel {\it b} the dashed black lines are
the empirical cuts used by \citet{gro06} to select the lowest
metallicity Seyfert 2s ($Z \sim 1 Z_\odot$). In both panels, the top regions
bordered by the dashed black curve/lines are where the low-metallicity Seyfert 2s are
located.
} \label{fig:metal} \end{figure*}

Figure~\ref{fig:lratio}{\it a} shows that both photoionization models can
reproduce the strong oxygen lines, since such line ratios are mainly
determined by the average ionization parameter. But the two begin to
disagree in Fig.~\ref{fig:lratio}{\it b}, where weaker lines like He\,{\sc
ii}\,$\lambda4686$ and \othree\,$\lambda4363$ are involved. This
discrepancy arises because of the different density structures of the
modeled clouds. 
In the dusty model (Model 1) the hydrogen density in a cloud varies from
a few hundreds to $\sim$1000 \cc. In this diagram, Model 1 cannot
provide a good fit to the data unless an extremely low metallicity
($\lesssim 0.2 Z_\odot$) is used, which nonetheless contradicts 
the best fit grids in Fig.~\ref{fig:lratio}{\it a} and Fig.~\ref{fig:metal} 
(\S~\ref{sec:metal}). 
However, the low-density MB component in the two-phase model can
easily reproduce both the high temperature ($\sim$ 15000 K) and the low
He\,{\sc ii} fluxes. Since most of the \othree\ is emitted by the MB
cloud, the influence from the IB cloud on temperature is negligible.
We emphasize that the high-density IB component is indispensible to
achieve the overall best fit, because the high \otwo\ fluxes cannot be
explained by the high-excitation MB component alone.  Note that since
the emission lines involved in these diagrams are fairly insensitive to
density effects, lowering density in the dusty model has the same effect
as increasing the ionization parameter, and hence would not improve the
fit to the data. 

The failure of Model 1 shows that these clouds are probably not as
dense as a giant molecular cloud (GMC). In fact, as argued by
\citet{sto02}, the general lack of correlation between the morphology of
the EELR and that of the stars in the host galaxy argues against them
being GMCs, since GMCs are supposedly dense enough that their
trajectories should follow those of the stars and not be
affected seriously by hydrodynamic interactions. 

The success of the simple two-phase model suggests that these clouds
consist of a general low-density medium together with a clumpy
distribution of much denser gas. This picture is also supported by the
amount of reddening and the density-sensitive line ratios observed in
these clouds (Table~\ref{tab:prop}). An intrinsic reddening of $A_V \sim
0.6$ mag is generally seen in such clouds, which implies a hydrogen
column density of $\sim 10^{21}$ cm$^{-2}$.  Assuming that the depth along
the line-of-sight
of the clouds is comparable to the resolved width of E1 (1\arcsec\ $\sim$
5 kpc), the resulting number density is $\sim$0.1 \cc.  But we
constantly measure a few hundreds \cc\ from either \otwo\ or [S\,{\sc
ii}] ratios. A natural solution to this apparent contradiction is that
clumpy dense clouds coexist with a much more diffuse medium. The former
produces most of the low-ionization emission lines such as the
\otwo\,$\lambda\lambda$3726,3729 and [S\,{\sc
ii}]\,$\lambda\lambda$6717,6731 doublets, leading to the measured high
densities; but the latter holds most of the mass. 
As pointed out by \citet{sto02}, the dense clouds need to be
continuously supplied, possibly through compression of the diffuse
medium by low-speed shocks, because (1) the two kinds of clouds are not
in pressure equilibrium since the temperatures of the two are both
around $10^4$ K, and (2) the denser clouds are geometrically thin
($\sim$0.1 pc, at least for the ionized part), thus cannot be stable on
a timescale longer than $10^4$ yr, given a typical sound speed of $\sim$
17 \kms. We will discuss the origin of the high-density medium in more
detail in \S~\ref{sec:highd}.

\subsection{Metallicity} \label{sec:metal}

Diagrams involving the metallicity sensitive [N\,{\sc
ii}]\,$\lambda6584$ line are shown in Figure~\ref{fig:metal}. 
The only clouds where we have spectral coverage reaching the
H$\alpha$-[N\,{\sc ii}] region are those observed with DEIMOS (N1 and
S1). Although the two clouds are about 40 kpc apart in projected
distance, their line ratios fall very close in Fig.~\ref{fig:metal},
suggesting similar metallicities. 
\citet{bor84} presented emission-line fluxes (their Table 2) of an E1
spectrum that covered the H$\alpha$-[N\,{\sc ii}] region.  After
dereddening, their data are consistent with the line ratios we measure
from the LRIS spectrum of E1. Because of their limited spectral
resolution, they only listed the total flux from H$\alpha$ + [N\,{\sc
ii}]\,$\lambda\lambda$6548,6584.  We roughly decomposed the H$\alpha$
and [N\,{\sc ii}]\,$\lambda6584$ fluxes by assuming the case B value of
H$\alpha$/H$\beta$ = 2.9 and a theoretical ratio of the [N\,{\sc ii}]
doublet of 3.0. The line ratios of this region are also plotted in
Fig.~\ref{fig:metal} (labeled as ``E1p'' to distinguish from the E1
region represented by the LRIS spectrum).

The fact that different photoionization models with a same
abundance\footnote{Note that the total gas metallicities in abundance sets $a$ and $b$ (\S~\ref{sec:ionization}) used in the two-phase model are both about 1/3 $Z^{'}_\odot$, and specifically, the nitrogen abundances are log(N/H) = $-4.42$ and $-4.45$,
respectively.} fall precisely onto one another in these plots shows that
such line ratios are almost entirely dependent on the abundance.
Therefore, the grids from the dusty model (Model 1) can be used to
measure the metallicities of the EELR clouds, even though it cannot reproduce some of the line ratios (refer to \S~\ref{sec:ionization}).

By interpolating the grids of Model 1 in both panels of
Fig.~\ref{fig:metal}, we estimated that 12 +
log(O/H) = 8.55, 8.58, 8.42 for S1, N1 \& E1p, respectively (Hereafter we will
refer to 12 + log(O/H) as (O/H)). Since the metallicity determination
also slightly depends on the density and Model 1 used a density of
$\sim$1000, we have corrected the above values using the densities
listed in Table~\ref{tab:prop} and the approximate (O/H) $\propto$
log($N_e$) scaling given in the Eqn. [4] of \citet{sto98}. The metallicities
after the correction are (O/H) = 8.62, 8.68, 8.47 for S1, N1 \& E1p, respectively.
The lower metallicity of E1p indicates that there might be some
metallicity variations within an EELR.

\citet{sto98} obtained two abundance calibrations for AGN NLRs from photoionization models. 
The calibrations involve similar line ratios as those in Fig.~\ref{fig:metal}, and were shown to yield consistent abundance values with those from 
H\,{\sc ii} regions for a sample of Seyferts. Therefore it is reassuring that the abundances obtained from these two calibrations are consistent with our simple interpolated results.
The average of the two abundance calibrations of \citet{sto98} (their
Eqs. [2], [3] and [4]) gives (O/H) $\simeq$ 8.64, 8.68 and 8.54 for S1,
N1 and E1p, respectively. 

To compare the data with the Seyferts at similar redshifts from
the Sloan Digital Sky Survey (SDSS), we delineated the regions where the
lowest metallicity Seyfert 2s were found in SDSS spectroscopic galaxy
sample \citep{gro06} in Fig.~\ref{fig:metal}. The regions are on the
left side of the main Seyfert branch in both panels.  It is clear that
the three EELR clouds around 4C\,37.43 all show significantly lower
metallicities than typical NLRs of Seyferts ((O/H) $\gtrsim 8.8$; \citealt{gro06}).

\section{Discussion} \label{sec:discussion}

\subsection{The Low Metallicity of 4C\,37.43} \label{sec:lowmetal}

Quasars are usually found in metal-rich environments ((O/H) $\gtrsim$
(O/H)$^{'}_\odot$ = 8.93) (\eg\ \citealt{ham02}). The metallicity does
not show a correlation with redshift, but seems to increase with black
hole (BH) masses \citep{war03}. The BH mass in 4C\,37.43 is about 10$^9$
M$_\odot$ \citep{lab06}, which predicts (O/H) = 9.8 \citep{war03}.
However, the low flux ratios of N\,{\sc v}\,$\lambda1240$/C\,{\sc
iv}\,$\lambda1549$ ($\sim$ 0.04) and N\,{\sc v}/He\,{\sc
ii}\,$\lambda1640$ ($\sim$ 0.24) observed in the broad-line region (BLR)
of this quasar \citep{kur02} both indicate a significantly lower
metallicity, (O/H) $\approx$ 8.4 (\ie\ about 20 times lower than a
typical quasar with $M_{BH}$ = 10$^9$ M$_\odot$) according to the
theoretical predictions of \citet{ham02}. Nevertheless, this BLR
metallicity is close to those measured in the surrounding EELR
(\S~\ref{sec:metal}). 

This low metallicity is unexpected for a quasar with a BH mass of  10$^9$
M$_\odot$ if (1) the bulge-mass---BH-mass and the bulge-mass---metallicity
correlations hold for this object and (2) the gas in the BLR and the EELR is from 
the ISM of the host galaxy of the quasar.  There is no indication that
the host galaxy of 4C\,37.43 is particularly low mass: it is easily visible
in ground-based and {\it HST} imaging \citep{sto02}, and, in fact, \citet{lab06}
find that the BH mass for 4C\,37.43 from the host galaxy luminosity
falls nicely between BH masses from the broad emission line widths
calculated under two different assumptions regarding the geometry of 
the BLR (see their Table 4).

Thus, whatever mechanism establishes the bulge-mass---BH-mass 
correlation likely was in operation in 4C\,37.43 long before the 
current episode of quasar activity, and the BH almost certainly
underwent most of its growth phase early on.  The low metallicity
of the gas in the BLR and EELR therefore indicates that this
gas has come from an external source, most likely a late-type
gas-rich galaxy that has recently merged.  If this is true, it provides
the first {\it direct} observational evidence that gas from a merger
can find its way down to the very center of the merger remnant,
to within fueling distance of the active nucleus, although there has
been a compelling statistical argument that this is the case for at
least some QSOs \citep{can01}.

\subsection{The Origin of the High-Density Medium} \label{sec:highd}

The viscosity of a fully ionized hydrogen plasma at $N_e$ = 1 \cc\ and $T
= 10^4$ K is $\mu \approx 9.5\times10^{-7}$ g s$^{-1}$ cm$^{-1}$
\citep{spi62}. Such a low viscosity implies a large Reynolds number if
the scale and the speed of the flow are large. In fact, for a typical EELR
cloud, R = $V m_p N l / \mu$ $\approx$ 10$^{11}$, given a flow velocity
$V$ = 200 \kms, a number density $N$ = 1 \cc\ and a typical length scale
$l$ = 1 kpc. This means that turbulent motions can be easily triggered
if the clouds are moving, converting part of its bulk kinetic energy
into turbulent energy.

The observed line-widths are around $\sigma \sim 50$ \kms, while the
sound speed in an ionized hydrogen plasma at $10^4$ K is only $c_s$
$\sim$ 17 \kms.  Shocks with a velocity $\sim$50 \kms\ must be common
inside the cloud. If the pre-shock gas is $10^4$ K, an isothermal shock
with a speed of 50 \kms\ is able to compress the gas by only a factor of
10 ($\rho_2/\rho_1 = M^2$, where $M$ is the Mach number). But if the
pre-shock gas is shielded from the ionizing source by the compressed
post-shock gas, then the temperature of the pre-shock gas is probably
$\sim$100 K, which corresponds to a sound speed of just $\sim$2 \kms.
The same shock speed thus could increase the gas density by over 600
times, sufficient to explain the high density contrast ($\sim$300)
between the two phases, as required by our best fit photoionization
model (\S~\ref{sec:ionization}).
The total mass in the dense medium can stay relatively stable as long as
the mass loss rate due to thermal expansion is comparable to the mass
accretion rate through shock compression.
We note that the amount of neutral gas due to this self-shielding should
be inappreciable compared to the total mass of the cloud, since the filling
factor of the dense material is known to be minuscule ($\phi \sim 10^{-5}$;
\citealt{sto02}).
Also, for shocks at such a low speed, the amount of UV photons produced
by the shock itself cannot significantly ionize the pre-shock neutral
gas ($U \approx 6\times10^{-5}$ for $V_S = 50$ \kms\ and $N = 1$ \cc;
\citealt{dop96}). 

Supersonic turbulence decays rapidly, typically over a
crossing timescale. However, since these clouds are large, the crossing
time is in fact on the same order-of-magnitude as the dynamical
timescale of the EELR, $\tau_{cross} \approx 20 (l/1$ kpc)/($v_t/50$ \kms) Myr;
so the picture of dense regions produced by shocks remains
consistent with the timescales involved.

\subsection{Mass of the Ionized Gas} \label{sec:mass}

The H$\beta$ luminosity of an ionized cloud is proportional to $N_e N_p
V$, where $V$ is the volume occupied by the emitting material. So, if we
assume $N_H$ = $N_e$ = $N_p$, then the total hydrogen mass can be
derived once we have an estimate of $N_e$: $M$ = $4\pi m_p f_{\rm
H\beta} d_L^2 / (\alpha_{\rm H\beta} N_e h\nu)$ = $3.2\times10^8$
$f_{\rm H\beta,-16}/N_{e,1}$ M$_{\odot}$, where $m_p$ is the proton
mass, $d_L$ the luminosity distance, $\alpha_{\rm H\beta}$ the effective
recombination coefficient of H$\beta$, $h\nu$ the energy of a H$\beta$
photon, $f_{\rm H\beta,-16}$ the H$\beta$ flux in units of $10^{-16}$
erg cm$^{-2}$ s$^{-1}$ and $N_{e,1}$ the electron density in units of 1
\cc. We assume a nominal density of 1 \cc\ because, as argued in
\S~\ref{sec:ionization}, most of the mass is in the low-density medium
with a density of $\sim$1 \cc\ even though the electron densities
measured from low-excitation lines are around hundreds \cc. The total
mass of the entire EELR is $\sim$ $3\times10^{10}$ M$_\odot$, given that
the total H$\beta$ flux is $1\times10^{-14}$ erg cm$^{-2}$ s$^{-1}$. 

One can also infer the cloud mass from the amount of dust observed and a
given dust-to-gas ratio. The dust column density is proportional to the
amount of reddening we see along the line-of-sight. Assuming the
Galactic dust-to-gas ratio, we obtain $M$ = $5.8\times10^{21}$ cm$^{-2}$
mag$^{-1}$ E$_{B-V} m_p d_A^2 \theta^2$ = $4.0\times10^8$ A$_V$
$\theta^2$ M$_{\odot}$, where A$_V$ is the intrinsic reddening in
magnitude, $\theta^2$ is the solid angle subtended by the cloud in
$\square\arcsec$. This result is consistent with what we obtained from
the H$\beta$ luminosity if $N_e$ $\sim$ 1 \cc, validating the existence of
the low-density medium. Taking this one step further, if we require the two
masses to be consistent, then it is easy to show that $N_e$ = 0.8
($f_{\rm H\beta,-16}$/$\theta^2$)/A$_V$ \cc. 
The H$\beta$ surface brightness of the EELR peaks at $\sim
5\times10^{-16}$ erg cm$^{-2}$ s$^{-1}$ arcsec$^{-2}$ for the
brightest regions and at $\sim 1\times10^{-16}$ erg cm$^{-2}$
s$^{-1}$ arcsec$^{-2}$ for more average clouds
(Fig.~\ref{fig:vfield}; given an \othree/H$\beta$ ratio of 9). If we
increase this value by a factor of 2 to account for the smearing due to
seeing, we end up with $N_e$ $\lesssim$ 10 \cc\ for the brightest
clouds and $N_e$ $\lesssim$ 2 \cc\ for the less luminous ones, given
A$_V$ = 0.8 mag.

The mass of the EELR is on the same order of magnitude as that of the
total ISM in the Galaxy, which implies that perhaps almost the entire ISM of the
quasar host galaxy (or, more likely, that of the merging partner) is
being ejected as well as having been almost fully ionized by the
quasar nucleus.

\subsection{The Driving Source of the Outflow} \label{sec:outflow}

The mass of the EELR, interpreted as an outflow, implies an enormous
amount of energy.  The bulk kinetic energy of the entire EELR is
approximately $E_{KE} = M V^2 / 2 = 6.2\times10^{57} M_{10} V_{250}^2$
ergs, where $M_{10}$ is the total mass in units of $10^{10}$ M$_{\odot}$
and $V_{250}$ = $V$/250 \kms.  We used 250 \kms\ as the nominal velocity
scale because most of the mass is in the brightest two concentrations,
which are both blueshifted by about 250 \kms.
The kinetic energy of the unresolved kinematic substructures
(``turbulent" energy) can be estimated from the measured line widths,
$E_{tur} = M \sigma^2 = 5.0\times10^{56} M_{10} \sigma_{50}^2$ ergs
($\sigma_{50} = \sigma/50$ \kms).
The emission-line gas also has a thermal energy of $E_{TH} = 1.5 M k T /
m_p = 2.5 \times 10^{55} M_{10} T_4$ ergs ($T_4 = T(K)/10^4 K$), which
is just about 0.4\% of the kinetic energy. 
Similarly, the total momentum is $p$ = $M V$ = $5\times10^{50} M_{10}
V_{250}$ dyne s.
Whatever the driving source of this outflow is, it must have deposited
this enormous amount of energy and momentum into the EELR in a short time
(a few $\times$10$^7$ yr, the dynamical timescale of the EELR). 
Such high input rates are consistent with an outflow driven by a quasar \citep{fu06}. 

Although in most cases the EELR morphology seems independent of the
structure of the radio source, the fact that both the occurrence and
luminosity of the EELR increase with the radio spectral index
\citep{sto87} suggests a link between the extended emission and the
radio jet. 
Recent simulations of jet-cloud interactions in elliptical galaxies show
that wide-solid-angle ``bubbles" could accompany the launching of the
jet (M. Dopita 2007, private communication). This is an intriguing
picture, because such bubbles provide a natural way to expel clouds from
the vicinity of the quasar to radii of a few tens of kpc
without resulting in any morphological similarities between the expelled
material and the jet, while also producing the observed connection
between radio outflows and EELRs.

This mechanism for globally expelling a quantity of gas comparable to the total
ISM of a reasonably massive galaxy is certainly of interest in the context of
current speculation regarding quasar winds as a means of initially establishing
the observed bulge-mass---black-hole-mass correlation (regardless of the extent
to which it has to be maintained by heating the surrounding medium via less violent
nuclear activity).  The main differences are that, (1) in 4C\,37.43, the black-hole has already 
achieved a mass of $\sim10^9$ M$_{\odot}$, presumably in a previous quasar phase, 
and it is currently ejecting rather low-metallicity gas that likely comes from a late-type, 
gas-rich merging companion; and (2) the ejection of the gas is a direct consequence
of large-scale shocks produced by the radio jet, rather than due to radiative coupling
of the quasar luminosity to the gas.

\section{Summary}

Our results suggest the following overall picture for the origin of the EELR around 4C\,37.43.
A large, late-type galaxy with a mass of a few $\times10^{10}$  M$_{\odot}$ of 
low-metallicity gas has recently merged with the gas-poor host galaxy of 4C\,37.43, which has a $\sim10^9$ M$_{\odot}$ BH at its center.  The low metallicity of the BLR gas indicates that gas from the merging companion has been 
driven to the center during the merger, triggering the quasar activity, including the production
of FR II radio jets.  The initiation of the jets also produces a wide-solid-angle blast wave
that sweeps most of the gas from the encounter out of the galaxy.  This gas is photoionized
by UV radiation from the quasar, and turbulent shocks produce high-density 
($\sim400$ cm$^{-3}$) filaments or sheets in the otherwise low-density ($\sim1$ cm$^{-3}$)
ionized medium.  The EELR will have a lifetime
% of $\sim2\times10^7$ years, 
on the order of 10 Myr, i.e., comparable
to that of the extended radio source.

In more detail, we can summarize the major conclusions of this paper as follows:
\begin{enumerate}
\item The EELR of 4C\,37.43 exhibits rather complex kinematics which cannot be 
explained globally by a simple dynamical model.
\item The \otwo\ or [S\,{\sc ii}] electron densities of the clouds range from 
600 \cc\ to less than 100 \cc. The R$_{\rm OIII}$ temperatures are mostly 
$\sim1.5\times10^4$ K. The cloud ($e$) having the lowest temperature ($T \sim 1.2\times10^4$ K)
also shows the lowest density, indicating a lower-than-average
pressure.
\item The spectra from the clouds are inconsistent with shock or ``shock +
precursor'' ionization models, but they are consistent with
photoionization by the quasar nucleus. 
\item The best-fit photoionization model requires a two-phase medium, 
consisting of a matter-bounded diffuse
component with a unity filling-factor ($N \sim 1$ \cc, $T \sim 15000$
K), in which are embedded small, dense clouds ($N \sim 400$ \cc, $T \sim
10^4$ K), which are likely constantly being regenerated through compression
of the diffuse medium by low-speed shocks. 
\item The metallicity of the EELR ((O/H) $\lesssim$ 8.7) is similar to 
that of the rare low-metallicity Seyferts \citep{gro06}, as implied by 
[N\,{\sc ii}] $\lambda$6584 line ratios and the overall best-fit photoionization 
model.  Previous results show that the BLR has a similarly low
metallicity, indicating a common (external) origin for the gas, which
also presumably fuels the current quasar activity.
\item The photoionization model gives a total mass for the ionized gas of
$3\times10^{10}$ M$_{\odot}$. The total kinetic energy implied by this
mass and the observed velocity field is $\sim2\times10^{58}$ ergs.   
\item The strong correlation of luminous EELRs with steep-spectrum radio-loud 
quasars, coupled with the general {\it lack} of significant correlation between 
the EELR and radio morphologies implies that the highly collimated radio jets 
are accompanied by a more nearly spherical blast wave. 

\item Since the mass of the ionized, apparently ejected gas in 4C\,37.43 is 
comparable to that of the ISM of a moderately large spiral, this object provides 
a local analog to (though likely not an example of) the hypothesized 
``quasar-mode'' ejection that may be instrumental in initially establishing 
the bulge-mass---black-hole-mass correlation at high redshifts.  
\end{enumerate}

\acknowledgments
We thank Tracy Beck and Gelys Trancho for helping with the GMOS
observations, Marc Kassis and Chuck Sorenson for supporting the DEIMOS
observations, and Mike Dopita, Lisa Kewley, and Brent Groves for helpful
discussions on the photoionization models. We thank Steve Rodney and
Ryan Foley for kindly providing the DEIMOS standard star observation
files.  We also thank the anonymous referee for a careful reading of the 
manuscript, for his or her concern that the paper be presented in a way that
would make it accessible to a wide audience, and for cogent comments to
help us clarify and improve the presentation.
This research has been partially supported by NSF grant AST03-07335.

%%%%%%%%%%%%%%%%% T A B L E S %%%%%%%%%%%%%%%%%%%
\clearpage
\begin{landscape}

\begin{deluxetable}{lccccccccc}
%\rotate
\tablewidth{0pt}
%\tabletypesize{\scriptsize}
\tablecaption{Properties of 4C\,37.43 EELR clouds
\label{tab:prop}}
\tablehead{
\colhead{Region}
& \colhead{$V$}
& \colhead{$\sigma$} 
& \colhead{$A_V$\tablenotemark{a}}
& \colhead{H$\beta \times 10^{17}$}
& \colhead{[O\,{\sc ii}]}
& \colhead{[S\,{\sc ii}]}
& \colhead{[O\,{\sc iii}]}
& \colhead{$N_e$\tablenotemark{b}}
& \colhead{$T_e$}
\\
\colhead{}
& \colhead{(\kms)}
& \colhead{(\kms)}
& \colhead{}
& \colhead{(erg cm$^{-2}$ s$^{-1}$)}
& \colhead{3726/3729}
& \colhead{6717/6731}
& \colhead{(4959+5007)/4363}
& \colhead{(\cc)}
& \colhead{(K)}
}
\startdata
  a  &  $-363$  &     \phn68  &  (0.700) &    \phn\phn$4.4\pm0.4$  &  $1.17\pm0.22$  &  \nodata        &   \nodata          &  $580\pm300$ &   \nodata               \\
  b  &   $-121$  &    133  &  (0.542) &     \phn\phn$6.9\pm0.3$  &  $1.08\pm0.10$  &  \nodata        &   \nodata          &  $460\pm130$ &   \nodata               \\
  c  &   $-150$  &     \phn79  &   0.540  &     \phn\phn$8.2\pm0.4$  &  $0.86\pm0.09$  &  \nodata        &      \phn$77\pm24$     &  $230\pm120$ & $14400^{+2900}_{-1500}$ \\
  d  &   \phn$-55$  &    116  &  (0.542) &     \phn\phn$5.9\pm0.4$  &  $1.07\pm0.24$  &  \nodata        &   \nodata          &  $450\pm310$ &   \nodata               \\
  e  &    \phn$-13$  &     \phn69  &   1.534  &   $164.6\pm2.2$  &  $0.68\pm0.05$  &  \nodata        &     $107\pm15$     &   $20\pm50$  & $12500^{+770\phn}_{-600\phn}$   \\
  f  &     \phn\phs$85$  &     \phn79  &   1.362  &    \phn$23.1\pm0.7$  &  $0.98\pm0.11$  &  \nodata        &      \phn$69\pm26$     &  $390\pm160$ & $15000^{+4200}_{-1900}$ \\
  g  &    \phs$167$  &    101  &  (0.700) &     \phn\phn$9.4\pm0.5$  &  $0.89\pm0.12$  &  \nodata        &   \nodata          &  $230\pm140$ &   \nodata               \\
  h  &    \phs$214$  &     \phn63  &   1.380  &    \phn$23.3\pm0.9$  &  $1.04\pm0.11$  &  \nodata        &      \phn$57\pm14$     &  $500\pm170$ & $16400^{+2600}_{-1600}$ \\
 E1  &   $-250$  &     \phn80  &   0.655  &   $317.2\pm3.2$  &  $0.97\pm0.03$  &  \nodata        &      $75\pm4$      &  $375\pm40$\phn  & $14500\pm400$           \\    
 S1  &    \phn$-73$  &    162  &   0.542  &     \phn\phn$6.8\pm0.4$  &  \nodata        &  $1.24\pm0.10$  &   \nodata          &  $190\pm120$ &   \nodata               \\
 N1  &    \phn$-59$  &    148  &   0.237  &    \phn$11.7\pm0.3$  &  \nodata        &  $1.46\pm0.11$  &      \phn$66\pm25$     &    $\leq60$  & $15300^{+4300}_{-2000}$ \\
\enddata
\tablenotetext{a}{Intrinsic reddening. The values in parentheses are not directly measured from the H$\gamma$/H$\beta$ ratio, since the H$\gamma$ lines in those
spectra are not well-detected. See \S~\ref{sec:ne_t} for details.}
\tablenotetext{b}{If there is a good measurement of the
temperature sensitive \othree\ intensity ratio, the electron density is derived
consistently with the R$_{\rm OIII}$ temperature. However, the quoted 1-$\sigma$ error
bars do not include the uncertainties of the other parameter.
In other cases, $N_e$ is derived assuming $T_e$ = $10^4$ K. }
\end{deluxetable}

%\clearpage
\begin{deluxetable}{lccccccccccc}
%\rotate
\tablewidth{0pt}
\tabletypesize{\tiny}
\tablecaption{Line Ratios of 4C\,37.43 EELR Clouds Relative to H$\beta$
\label{tab:liner}}
\tablehead{
\colhead{Region}
& \colhead{[Ne\,{\sc V}]\,$\lambda3426$} 
& \colhead{[O\,{\sc II}]\,$\lambda3726$}
& \colhead{[O\,{\sc II}]\,$\lambda3729$}
& \colhead{[Ne\,{\sc III}]\,$\lambda3869$}
& \colhead{[O\,{\sc III}]\,$\lambda4363$}
& \colhead{[He\,{\sc II}]\,$\lambda4686$}
& \colhead{[O\,{\sc III}]\,$\lambda5007$} & \colhead{H$\alpha$}
& \colhead{[N\,{\sc II}]\,$\lambda6584$} & \colhead{[S\,{\sc II}]\,$\lambda6716$}
& \colhead{[S\,{\sc II}]\,$\lambda6731$}
}
\startdata
  a  & $<1.51$        &  $1.53\pm0.18$  &  $1.31\pm0.19$  &  $0.85\pm0.27$  &  $<0.41$        &  $<0.29$        &   \phn$9.30\pm0.13$  &  \nodata        &  \nodata        &  \nodata        &  \nodata        \\ 
  b  & $<0.98$        &  $2.35\pm0.15$  &  $2.17\pm0.14$  &  $1.04\pm0.14$  &  $<0.28$        &  $<0.18$        &    \phn$6.88\pm0.07$  &  \nodata        &  \nodata        &  \nodata        &  \nodata        \\ 
  c  & $<0.61$        &  $2.10\pm0.09$  &  $2.43\pm0.23$  &  $1.27\pm0.11$  &  $0.15\pm0.05$  &  $<0.13$        &    \phn$8.50\pm0.13$  &  \nodata        &  \nodata        &  \nodata        &  \nodata        \\ 
  d  & $<1.03$        &  $1.88\pm0.28$  &  $1.76\pm0.29$  &  $0.97\pm0.13$  &  $<0.30$        &  $<0.20$        &    \phn$8.21\pm0.08$  &  \nodata        &  \nodata        &  \nodata        &  \nodata        \\ 
  e  & $1.85\pm0.08$  &  $1.07\pm0.07$  &  $1.57\pm0.07$  &  $1.14\pm0.05$  &  $0.15\pm0.02$  &  $0.33\pm0.01$  &  $12.01\pm0.01$  &  \nodata        &  \nodata        &  \nodata        &  \nodata        \\ 
  f  & $<0.69$        &  $2.04\pm0.21$  &  $2.09\pm0.08$  &  $0.98\pm0.09$  &  $0.14\pm0.05$  &  $0.18\pm0.03$  &    \phn$7.37\pm0.11$  &  \nodata        &  \nodata        &  \nodata        &  \nodata        \\ 
  g  & $<0.62$        &  $1.67\pm0.20$  &  $1.89\pm0.11$  &  $0.80\pm0.12$  &  $<0.23$        &  $0.21\pm0.04$  &    \phn$9.73\pm0.10$  &  \nodata        &  \nodata        &  \nodata        &  \nodata        \\ 
  h  & $1.00\pm0.22$  &  $1.59\pm0.12$  &  $1.53\pm0.12$  &  $1.46\pm0.11$  &  $0.19\pm0.05$  &  $0.19\pm0.03$  &    \phn$8.30\pm0.03$  &  \nodata        &  \nodata        &  \nodata        &  \nodata        \\ 
 E1  & $0.36\pm0.02$  &  $1.42\pm0.03$  &	 $1.46\pm0.03$  &  $0.96\pm0.02$  &  $0.19\pm0.02$  &  $0.20\pm0.02$  &  $10.50\pm0.06$  &  \nodata        &  \nodata        &  \nodata        &  \nodata        \\
 S1  & $<0.38$        & \multicolumn{2}{c}{$4.42\pm0.10$\tablenotemark{a}}    &  $0.73\pm0.07$  &        $<0.17$  &  $0.22\pm0.05$  &    \phn$8.74\pm0.07$  &  $3.16\pm0.03$  &  $1.50\pm0.03$  &  $0.55\pm0.03$  &  $0.44\pm0.03$  \\
 N1  & $<0.14$        & \multicolumn{2}{c}{$3.53\pm0.05$\tablenotemark{a}}    &  $0.81\pm0.04$  &  $0.16\pm0.06$  &  $0.15\pm0.03$  &    \phn$8.01\pm0.04$  &  $3.16\pm0.03$  &  $1.56\pm0.03$  &  $0.64\pm0.03$  &  $0.44\pm0.03$  \\
\enddata
\tablenotetext{a}{These values are total fluxes in the
\otwo\ doublet, since the doublets are unresolved in these spectra.}
\end{deluxetable}
\clearpage
\end{landscape}

\clearpage
\begin{deluxetable}{lccccc}
\tablewidth{0pt}
\tabletypesize{\scriptsize}
\tablecaption{Abundances of key elements \label{tab:abund}}
\tablehead{
\colhead{Set$^1$} & \colhead{H} & \colhead{He} & \colhead{N} & \colhead{O} & \colhead{Ne}
}
\startdata
a & 0.00 & $-0.99$ & $-4.42$ & $-3.53$ & $-3.92$ \\ 
b & 0.00 & $-1.01$  & $-4.45$ & $-3.57$ & $-4.41$ \\
\enddata
\tablenotetext{1}{All abundances are logarithmic with respect to hydrogen.}
\end{deluxetable}

%%%%%%%%%%%%%%%%% F I G U R E S %%%%%%%%%%%%%%%%%%%
%\clearpage

\clearpage

\clearpage

\clearpage

\clearpage

\clearpage


\begin{thebibliography}

\bibitem[Allington-Smith et al.(2002)]{all02} 
Allington-Smith, J., et al.\ 2002, \pasp, 114, 892 

\bibitem[Anders \& Grevesse(1989)]{and89} Anders, E., \& 
Grevesse, N.\ 1989, \gca, 53, 197 

\bibitem[Binette et al.(1996)]{bin96} Binette, L., Wilson, 
A.~S., \& Storchi-Bergmann, T.\ 1996, \aap, 312, 365 

\bibitem[Boroson \& Oke(1984)]{bor84} Boroson, T. A., \& Oke, J. B.
1984, \apj, 281, 535

\bibitem[Boroson et al.(1985)]{bor85} Boroson, T.~A., 
Persson, S.~E., \& Oke, J.~B.\ 1985, \apj, 293, 120

\bibitem[Canalizo \& Stockton(2001)]{can01} Canalizo, G., \& Stockton, A.
   2001, \apj, 555, 719

\bibitem[Cappellari \& Copin(2003)]{cap03} Cappellari, M., \& 
Copin, Y.\ 2003, \mnras, 342, 345 

\bibitem[Cardelli et al.(1989)]{car89} Cardelli, J.~A., Clayton, G.~C.,
\& Mathis, J.~S.\ 1989, \apj, 345, 245 

\bibitem[Crawford et al.(1988)]{cra88} Crawford, C. S., Fabian, A. C., \& 
   Johnstone, R. M. 1988, \mnras, 235, 183

\bibitem[Crawford \& Vanderriest(2000)]{cra00} Crawford, 
C.~S., \& Vanderriest, C.\ 2000, \mnras, 315, 433 

\bibitem[Di Matteo et al.(2005)]{diM05} Di Matteo, T., 
Springel, V., \& Hernquist, L.\ 2005, \nat, 433, 604

\bibitem[Dopita \& Sutherland(1996)]{dop96} Dopita, M. A., \&
Sutherland, R. S.  1996, \apjs, 102, 161

\bibitem[Durret et al.(1994)]{dur94} Durret, F., Pecontal, 
E., Petitjean, P., \& Bergeron, J.\ 1994, \aap, 291, 392

\bibitem[Faber et al.(2003)]{fab03} Faber, S.~M., et al.\ 
2003, \procspie, 4841, 1657

\bibitem[Fabian et al.(1987)]{fab87} Fabian, A.~C., Crawford, C.~S.,
Johnstone, R.~M., \& Thomas, P.~A.\ 1987, MNRAS, 228, 963 

\bibitem[Ferland \& Osterbrock(1986)]{fer86} Ferland, G.~J., \& Osterbrock,
D.~E.\ 1986, \apj, 300, 658

\bibitem[Friaca \& Terlevich(1998)]{fri98} Friaca, A.~C.~S., 
\& Terlevich, R.~J.\ 1998, \mnras, 298, 399 

\bibitem[Fu \& Stockton(2006)]{fu06} Fu, H., \& Stockton, 
A.\ 2006, \apj, 650, 80

\bibitem[Gloeckler \& Geiss(2007)]{glo07} Gloeckler, G., Geiss, J.\ 2007, Space Science Reviews, in press.

\bibitem[Groves et al.(2004)]{gro04} Groves, B.~A., Dopita, M.~A., \&
Sutherland, R.~S.\ 2004, \apjs, 153, 9

\bibitem[Groves et al.(2006)]{gro06} Groves, B.~A., Heckman, 
T.~M., \& Kauffmann, G.\ 2006, \mnras, 371, 1559

\bibitem[Hamann et al.(2002)]{ham02} Hamann, F., Korista, 
K.~T., Ferland, G.~J., Warner, C., \& Baldwin, J.\ 2002, \apj, 564, 592 

\bibitem[Hook et al.(1994)]{hoo94} Hook, R., Lucy, L., 
Stockton, A., \& Ridgway, S.\ 1994, Space Telescope European Coordinating 
Facility Newsletter, Volume 21, p.16, 21, 16

\bibitem[Hook et al.(2004)]{hoo04} Hook, I.~M., J{\o}rgensen, 
I., Allington-Smith, J.~R., Davies, R.~L., Metcalfe, N., Murowinski, R.~G., 
\& Crampton, D.\ 2004, \pasp, 116, 425

\bibitem[Hopkins et al.(2006)]{hop06} Hopkins, P.~F., Hernquist, L., Cox, T.~J., 
   Di Matteo, T., Robertson, B., \& Springel, V. 2006, \apjs, 163, 1

\bibitem[Kewley \& Dopita(2002)]{kew02} Kewley, L.~J., \& 
Dopita, M.~A.\ 2002, \apjs, 142, 35

\bibitem[Kuraszkiewicz et al.(2002)]{kur02} Kuraszkiewicz, 
J.~K., Green, P.~J., Forster, K., Aldcroft, T.~L., Evans, I.~N., \& 
Koratkar, A.\ 2002, \apjs, 143, 257 

\bibitem[Labita et al.(2006)]{lab06} Labita, M., Treves, A., 
Falomo, R., \& Uslenghi, M.\ 2006, \mnras, 373, 551

\bibitem[Miller et al.(1993)]{mil93} Miller, P., Rawlings, 
S., \& Saunders, R.\ 1993, \mnras, 263, 425 

\bibitem[Oke et al.(1995)]{oke95} Oke, J.~B., et al.\ 1995, 
\pasp, 107, 375

\bibitem[Osterbrock(1989)]{ost89} Osterbrock, D. E. 1989, Astrophysics
of Gaseous Nebulae and Active Galactic Nuclei, Mill Valley, California:
University Science Books

\bibitem[Schlegel et al.(1998)]{sch98} Schlegel, D.~J., Finkbeiner,
D.~P., \& Davis, M.\ 1998, \apj, 500, 525 

\bibitem[Spitzer(1962)]{spi62} Spitzer, L.\ 1962, Physics of 
Fully Ionized Gases, New York: Interscience (2nd edition), 1962

\bibitem[Stockton(1976)]{sto76} Stockton, A.\ 1976, \apjl, 
205, L113

\bibitem[Stockton et al.(2006a)]{sto06} Stockton, A., Fu, H., Henry,
J.~P., \& Canalizo, G., 2006a, \apj, 638, 635

\bibitem[Stockton et al.(2006b)]{sto06b} Stockton, A., Fu, H., 
\& Canalizo, G.\ 2006b, New Astronomy Review, 50, 694 

\bibitem[Stockton \& MacKenty(1987)]{sto87} Stockton, A., \& MacKenty,
J. W. 1987, \apj, 316, 584

\bibitem[Stockton et al.(2002)]{sto02} Stockton, A., MacKenty, J. W.,
Hu, E. M., \& Kim, T.-S. 2002, \apj, 572, 735

\bibitem[Storchi-Bergmann et al.(1998)]{sto98} 
Storchi-Bergmann, T., Schmitt, H.~R., Calzetti, D., \& Kinney, A.~L.\ 1998, 
\aj, 115, 909 

\bibitem[Warner et al.(2003)]{war03} Warner, C., Hamann, F., 
\& Dietrich, M.\ 2003, \apj, 596, 72 

\end{thebibliography}
\end{document}